# Estimation of *In-Vitro* Free Radical Scavenging and Cytotoxic Activities of Thiocarbohydrazones Derivatives

**Qurat-ul-Ain[1,2*], Munira Taj Muhammad[2], Khalid Mohammed Khan[1,2,3], and M. Iqbal Choudhary[1,2,4]**

[1]*Dr. Panjwani Center for Molecular Medicine and Drug Research, International Center for Chemical and Biological sciences, University of Karachi-75270, Pakistan*

[2]*H. E. J. Research Institute of Chemistry, International Center for Chemical and Biological Sciences, University of Karachi, Karachi-75270, Pakistan*

[3]*Department of Clinical Pharmacy, Institute for Research and Medical Consultations (IRMC), Imam Abdulrahman Bin Faisal University, P.O. Box 31441, Dammam, Saudi Arabia*

[4]*Department of Biochemistry, Faculty of Science, King Abdulaziz University, Jeddah-21412, Saudi Arabia.*

***Corresponding author:** Qurat-ul-Ain, Dr. Panjwani Center for Molecular Medicine and Drug Research, International Center for Chemical and Biological sciences, University of Karachi-75270, Pakistan.

E-mail: Quratulain@iccs.edu

Foundation Project: Supported by Prof. Salim-uzz-Zaman Scholarship from HEJ Research Institute of Chemistry, ICCBS and Higher Education Commission of Pakistan

## ABSTRACT

**Objective:**

To investigate free radical scavenging potential of Thiocarbohydrazones derivatives for the discovery of antioxidant compounds of synthetic origin.

**Method:**

Eighteen Thiocarbohydrazones derivatives were screened for antioxidant activities by using *in vitro* 2,2-diphenyl-1-picrylhydrazyl (DPPH) inhibition assay. Cytotoxicity all derivatives were evaluated using *invitro* 3-(4,5-dimethylthiazol-2-yl)-2,5-diphenyltetrazolium bromide (MTT) assay on mouse fibroblast 3T3 cell line.

**Results:**

Compounds (**1-9**) and (**13-18**) have shown significant radical scavenging activities with $IC_{50}$ values in the range of 36 to 103 µM. Among all of the tested derivative's compound **10** was found to be the most potent inhibitor of the series with ($IC_{50}$ = 24.2 ± 0.12 µM) as compared with the standard drug butylated hydroxytoluene (BHT) ($IC_{50}$ = 128.8 ± 2.1 µM). Some other compounds of the series like compounds **12** exhibited week inhibitory activity with $IC_{50}$ value 208.5 ± 3.93 µM as compared with the standard drug BHT. In addition in an *in vitro* cell cytotoxicity assay selected analogues **1**, **2, 4, 5, 6, 7, 10, 11, 13, 16** and **17** were found to be non- cytotoxic against fibroblast 3T3 cell line.

**Conclusion:**

Taken together, these findings all compound **1, 2, 4, 5, 6, 7, 10, 11, 13, 16** and **17** of this series could be further evaluated in *in vivo* model of oxidative stress to develop as a therapeutic agent against oxidative stress related disorders due to exhibiting excellent free radical scavenging properties and having their non cytotoxic nature against normal cells.

**Key words:** Antioxidants, Radical Scavengers, DPPH, Thiocarbohydrazones, Cytotoxicity

Abbreviations: MTT: 3-(4,5-dimethylthiazol-2-yl)-2,5-diphenyl tetrazolium bromide

## 1. Introduction

Free radicals are short existing highly reactive molecular species containing unpaired electron that can potentially oxidized all major biological macromolecules within the cells and generate oxidative stress (OS) [1]. Free radicals include reactive oxygen species (ROS), reactive nitrogen species (RNS) and reactive sulfur species (RSS) [2]. Different types of ROS include Superoxide ($O_2$), Hydrogen peroxide ($H_2O_2$), Hydroxyl radical (HO), Peroxyl radical ($RO_2$), Alkoxyl radical (RO), Hydroperoxyl radical ($HO_2$), Singlet oxygen ($1O_2$) and Ozone ($O_3$) [3]. In the literature these molecules are reported as highly reactive in nature that physiologically produced in human body by the oxygen during various metabolic pathways [4]. Moreover, ROS can cause extensive damage to cells and tissues, during infections, cardiovascular disease, neurodegenerative diseases, and aging and also have been reported to be involved in both etiology and progression of several types of cancers [4].

Mammalian cells possess elaborate inherent defense antioxidant system in the form of various antioxidant enzymes for the detoxification of free radicals including the dismutation of superoxide to hydrogen peroxide and oxygen *via* superoxide dismutase (SOD), conversion of $H_2O_2$ into water and oxygen by catalase (CAT), and enzymatic reduction of toxic peroxides by glutathione peroxidase (GPX) [5]. In addition to antioxidant enzymes, several small-molecular antioxidants play important roles in the antioxidant defense systems. These can be divided into compounds i) obtained from diet such as glutathione, bilirubin, and melatonin, ii) vitamins such as α-tocopherol, beta-carotene, and ascorbic acid, and iii) micronutrient elements such as zinc and selenium [6]. Reduced expression of antioxidant enzymes and molecules have been reported in various cancers with higher levels of free radicals within the cancerous cells. Hence, scavenging of these free radicals could serve as an excellent approach to develop anticancer therapy [7, 8].

Antioxidants are the molecules of synthetic or plant sources that can reverse free radical induced damage either by scavenging free radicals, upregulating antioxidant enzyme, or by downregulating free radical inducing enzymes [9]. Many synthetic antioxidants have been reported for example resveratrol that upregulates the endogenous antioxidant systems, such as the SOD enzymes, in endothelial cells and cardiac myoblasts, and further decreases ROS production by scavenging free radicals [10]. In addition, another antioxidant anthocyanin was reported to increase SOD, CAT, and GPx, thereby found to prevent eye diseases such as age-related macular degeneration (AMD) in human retinal capillary endothelial cells [11]. Another study anthocyanins were found to ameliorate human retinal capillary endothelial function and revert the progression of diabetic retinopathy by decreasing ROS and increasing CAT and SOD activity [12]. Furthermore, berberine is a well-identified Chinese herbal medicine, that is present in the roots, rhizomes, and stem bulk of plants and has been observed to have significant antioxidant activity *in vitro* and *in vivo* models of oxidative stress and extensively used in the treatment of a wide range of inflammatory diseases [13-17]. Therefore, many such molecules of therapeutic potential belonging to plant source or synthesized in the laboratory have been reported so far in literature that have been used to reverse free radical induced damage *in vitro* and in *vivo* model systems and several are being testing in clinical trials to reverse the damage.

Thiocarbohydrazones (TCHs) are the homologue of thiosemicarbazones (TSCs), a class of compounds whose transition metal complexes have been extensively studied for their antimicrobial, antifungal, anti-cancer, antituberculosis, anti-inflammatory, anticonvulsant, antihypertensive, local anesthetic, antimycobacterial, antiviral, cytotoxic as well as antioxidative activities [18-20]. The attractiveness of TCHs for developing new metal-based drugs lies primarily in the presence of extra metal binding motif so they can form both mononuclear and binuclear complexes upon metal binding to a greater extent than TSCs [21]. In this paper our research group is reporting s *in vitro* free radical scavenging and cytotoxic activities of some new TSCs class derivatives. We have used DPPH and MTT method for evaluating free radical scavenging and cytotoxic activities of compounds of synthetic origin. DPPH assay exploits deep violet DPPH radical for evaluation free radical scavenging potential and MTT assay based on reduction of MMT by live cell mitochondrial dehydrogenase and considered as the best procedures for discovery of non-cytotoxic potential radical scavengers.

## 2. Material and Methods

### *2.1. Chemicals and Reagents*

All TCHs derivatives were obtained from the in-house Molecular Bank facility of the Dr. Panjwani Center for Molecular Medicine and Drug Research, International Center for Chemical and Biological Sciences, University of Karachi, Pakistan. DPPH was purchased from Sigma Aldrich (Germany). Ethanol and dimethyl sulfoxide (DMSO) (reagent grade) were purchased from Sigma Aldrich (USA). Standard compounds, *i.e.*, butylated hydroxytoluene (BHT) was purchased from Sigma Aldrich (Germany).

### *2.2. 2,2-diphenyl-1-picrylhydrazyl (DPPH) Inhibition Assay*

To carried out DPPH assay 5 µl of 500 µM solution of test compound was mixed with 95 µl of 300 µM freshly prepared ethanolic DPPH solution and then reaction was allowed to progress for 30 minutes at 37 ºC as previously described [22,23]. Absorbance was recorded by multiple reader (Spectra Max 340, Molecular Devices, CA, USA) at 517 nm. Upon reduction, the color of the solution fades away (Violet to pale yellow). Compounds scavenge 50% or more DPPH radical considered as active compounds that were further proceeded to measure their $IC_{50}$ values. To calculate $IC_{50}$ of each compound were diluted from 500 µM to 125 µM by two-fold dilution methods each concentration of test compound was treated with 95 µl of 300 µM prepared ethanolic DPPH solution for 30 minutes at 37 ºC and absorbance was recorded at 517 nm. The $IC_{50}$ values of compounds were calculated by using the EZ-Fit Enzyme kinetics software program (Perrella Scientific Inc., Amherst, MA, USA). As a standard compound we used BHT.

### *2.3. 3-(4,5-dimethylthiazol-2-yl)-2,5-diphenyl tetrazolium bromide (MTT) Assay*

Succinate dehydrogenase is an enzyme, which helps in the reduction of MTT dye into formazan salt in Mitochondria of the live cells [24]. The reduction of MTT dye is based on the metabolic activity of viable cells (Scheme-2). Compounds that are least cytotoxic showed more reduction of MTT dye, and thus higher color formation [25].

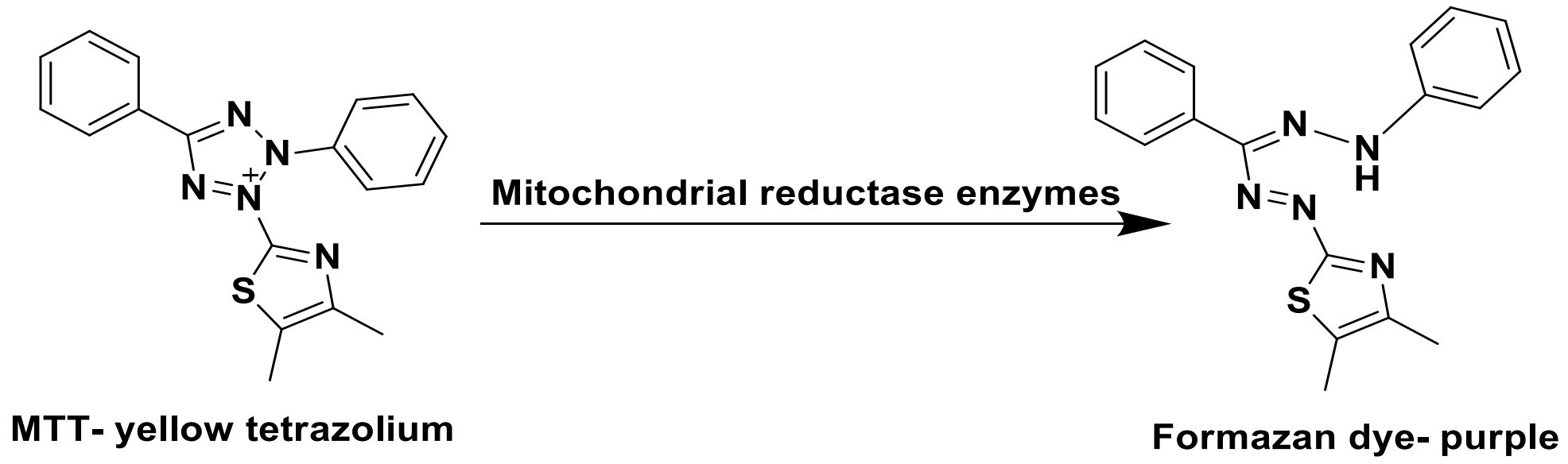

**Scheme-1**: **Reaction of MTT assay**

The assay was initiated with the addition of media (100 µl) with $3x10^5$ cells, in 96-well plate. Cells were allowed to attached to the bottom of the plate. Media was changed after 24 hours with 180 µl of fresh media then 20 µl of different concentrations (1 to 30 µM) of test compound was added in the plate. The plate was placed in carbon dioxide incubator (5%) with 90% humidity at 37 ºC for 48 hours. After incubation, MTT dye (200 µl) was added, and the plate was then incubated for 4 hours. Living cells crystalized with formazan, which were then dissolved by adding DMSO (100 µl). Absorbance was recorded at 540 nm. Estimation of $IC_{50}$ values were carried *via* EZ-Fit software. Cyclohexamide was used as standard comound.

**Dulbecco's modified eagle's medium (DMEM) was used for culturing 3T3 cells**

↓

**Add fetal bovine serum (5%) + Streptomycin (100 µg/ml) in 25 $cm^3$ flask + Penicillin (100 iU/ml)**

↓

**Overnight incubate at 37ºC, and separate media after incubation**

↓

**After 72 hours, add MTT (2 mg/ml) + 100 µl of DMSO**

↓

**50% inhibition of 3T3 cells used for cytotoxic evaluation**

**Scheme-2: procedure of MTT assay**

### *2.1. Statistical Analysis*

Each sample was tested in triplicate Statistical analysis for both MTT and DPPH assay were performed through EZ Fit software. Values of all parameters are expressed as mean ± SD of three independent measurements.

## 3. Results and Discussion

### *3.1. Chemistry and Synthesis*

All TCHs derivatives (**1-18**) were synthesized from commercially available thiocarbohydrazide by condensing with different commercially available aldehydes. The reaction was performed in ethanol in the presence of acetic acid (0.5 ml), the reaction mixture was refluxed for 2-3 hrs. [26].

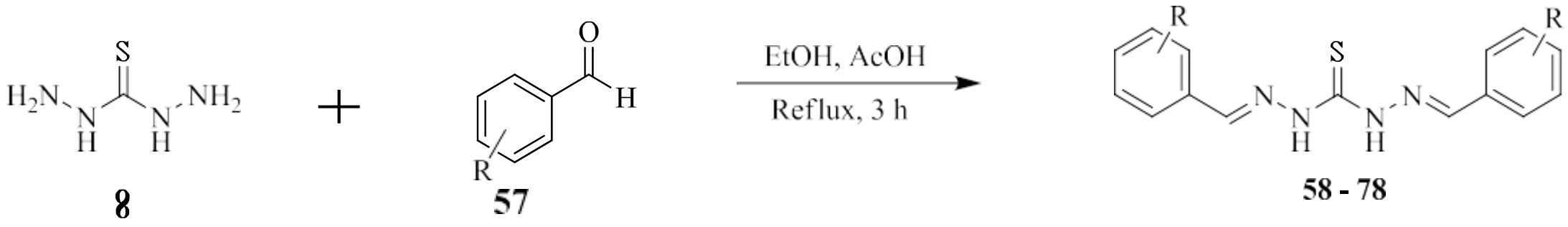

**Scheme-3**: **Synthesis of Thiocarbohydrazone (TCHs) (1-18)**

*3.2. Free Radical Scavenging Activity Against DPPH Radical*

Therefore, THCs **1-18** derivatives were evaluated for their free radical scavenging and cell cytotoxic activities. All THCs **(1-18)** were found to scavenge DPPH Radical with concentration dependent manner as presented in Figure1. Their radical scavenging activities ranged from 81.407% to 96.48% and the $IC_{50}$ values 40.41 ± 1.78 to 109. ± 0.98 µM compared to the standard BHT (--% RSA and 128.83 ± 2.1 µM $IC_{50}$) as presented in Table- 2. Compound **13** was found to be the most potent radical scavenger of the series with 96.48% radical scavenging potential and the corresponding $IC_{50}$ value 109. ± 0.98 µM. On contrary compound **4** was found to be the less potent scavenger of the series with 81.407% radical scavenging potential and the corresponding $IC_{50}$ value 40.41. ± 1.78 µM.

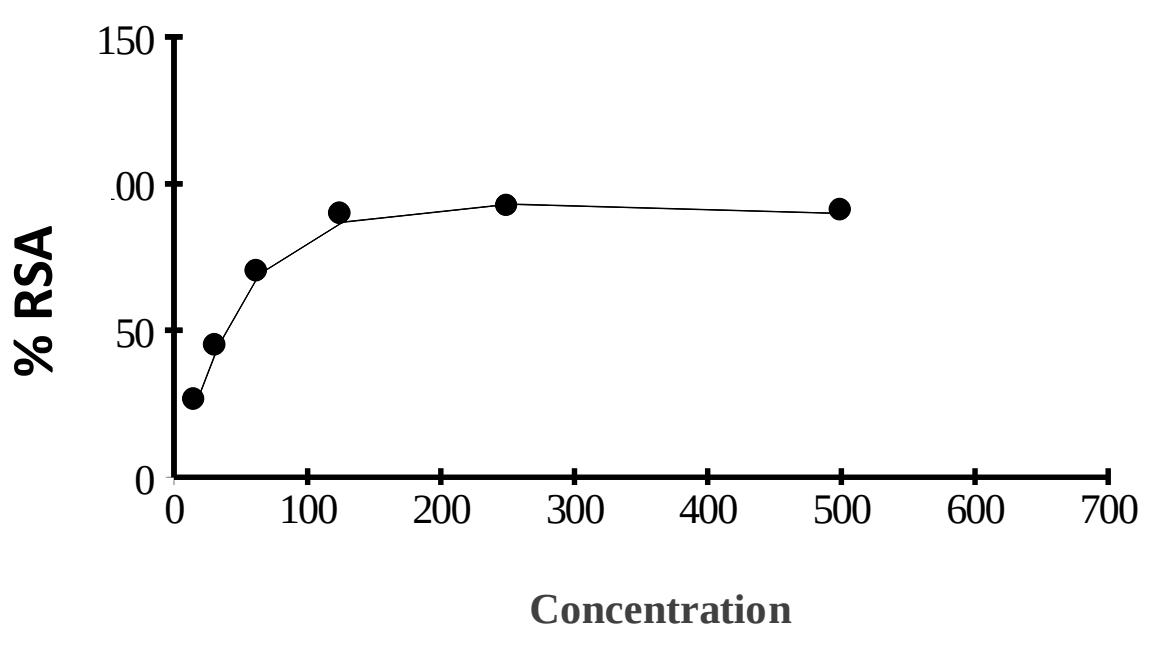

**Comp 1**

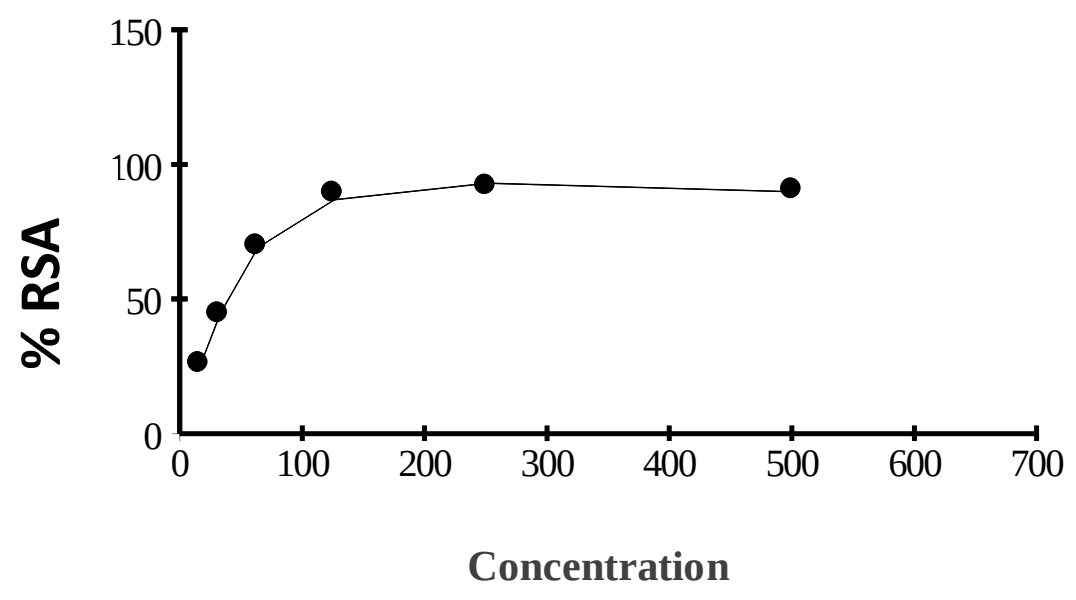

**Comp 1 5**

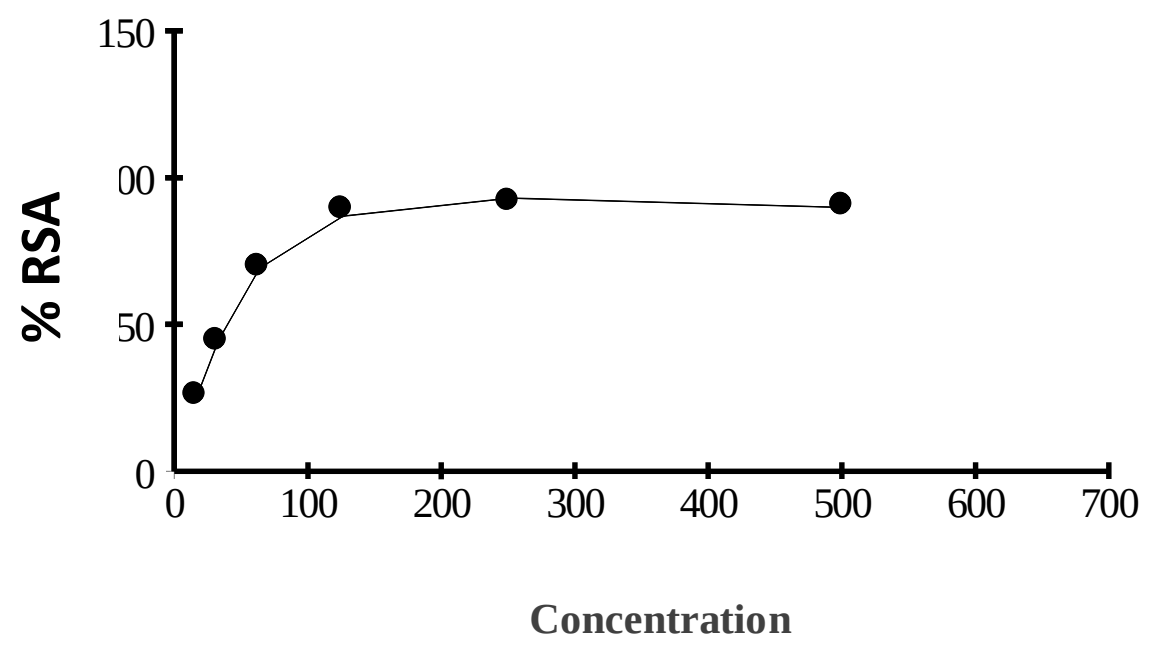

**Comp 4**

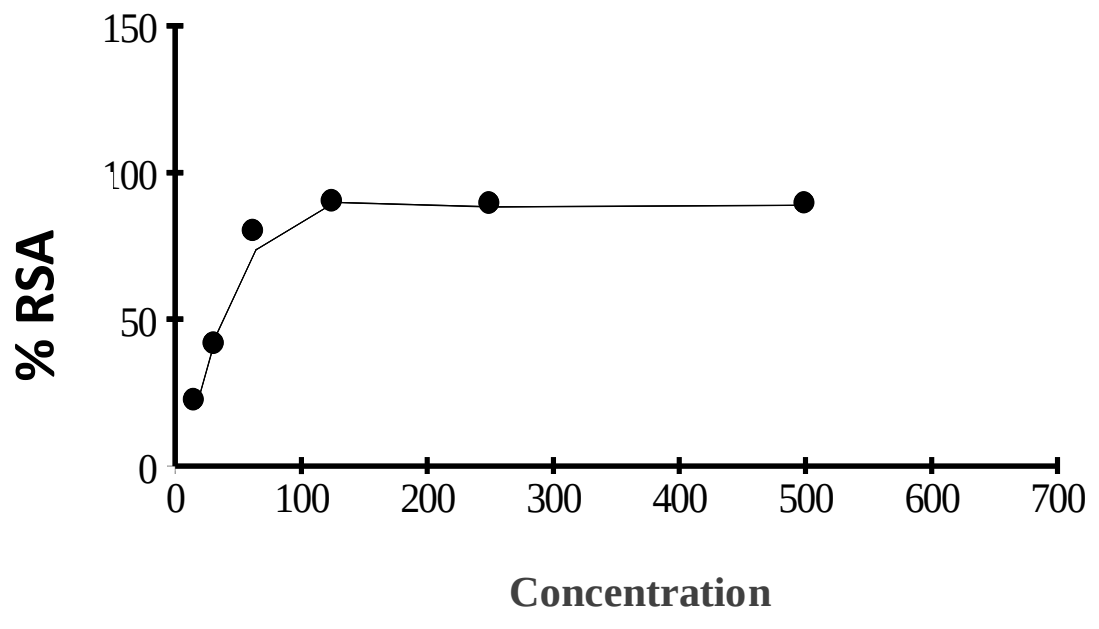

**Comp 9**

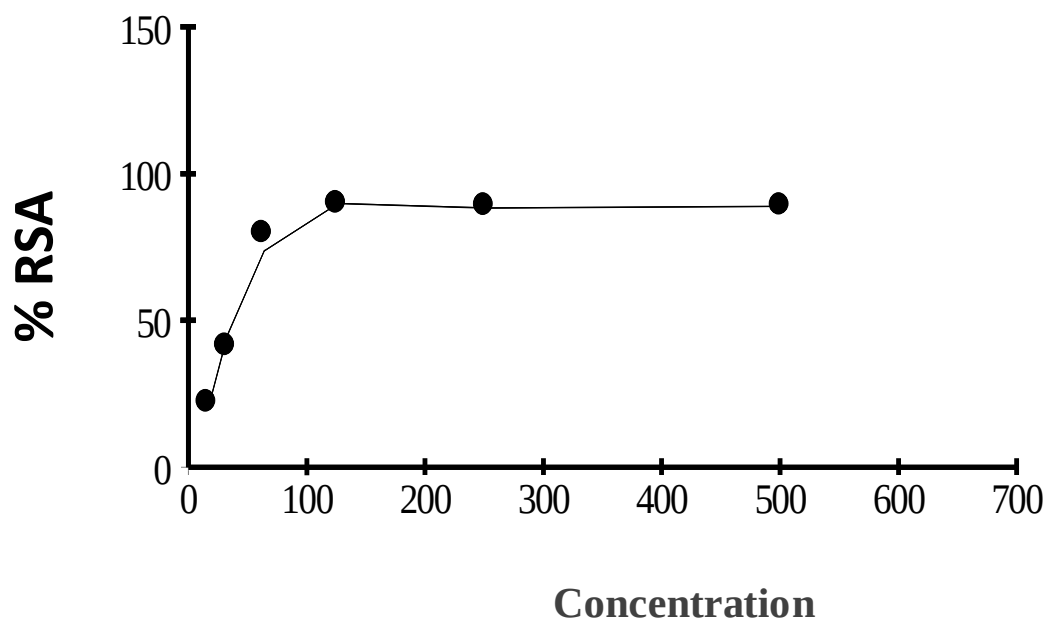

**Comp 1 7**

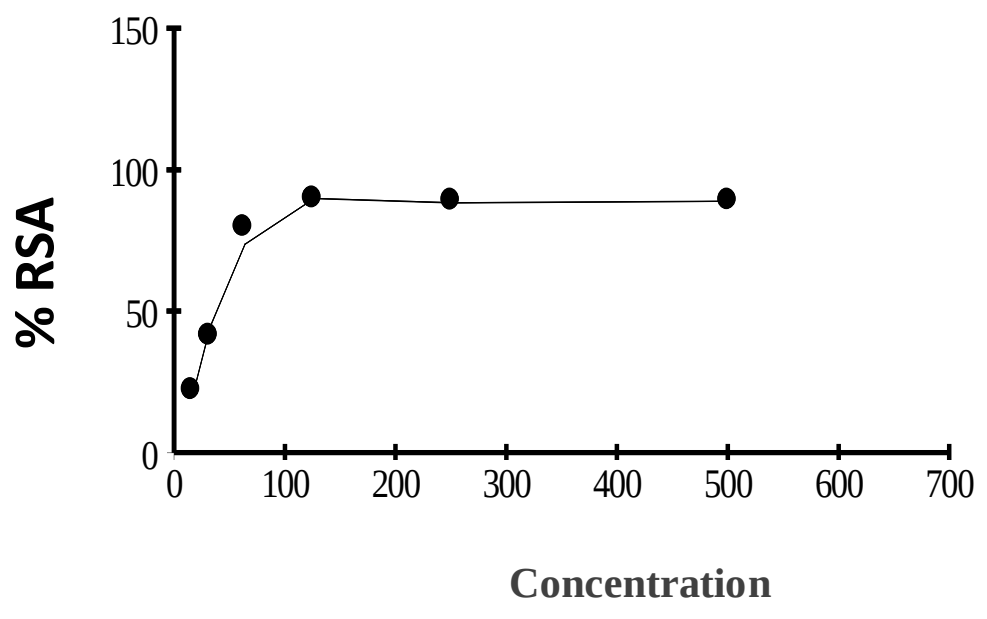

**Comp 1 0**

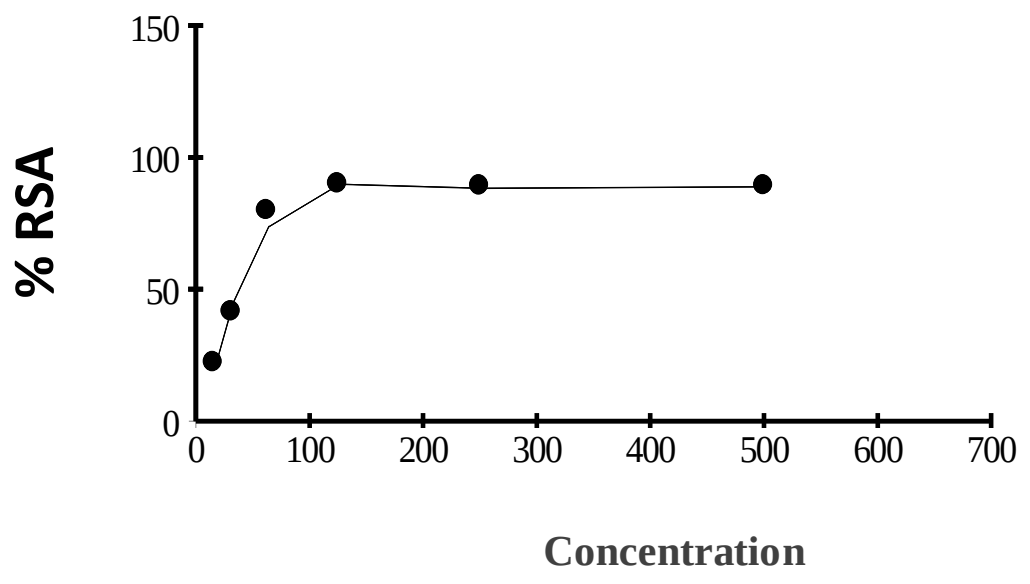

**Comp 5**

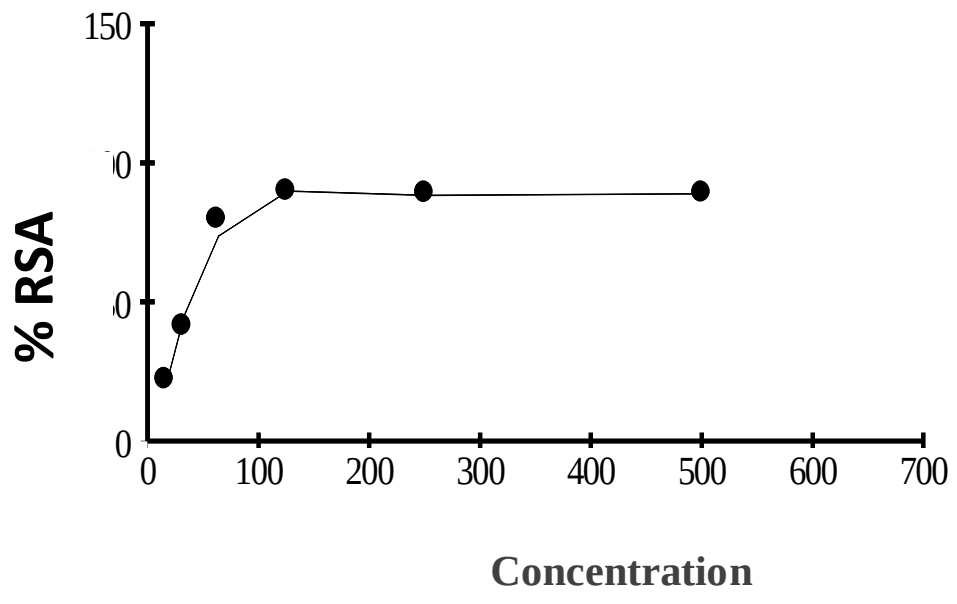

**Comp 1 6**

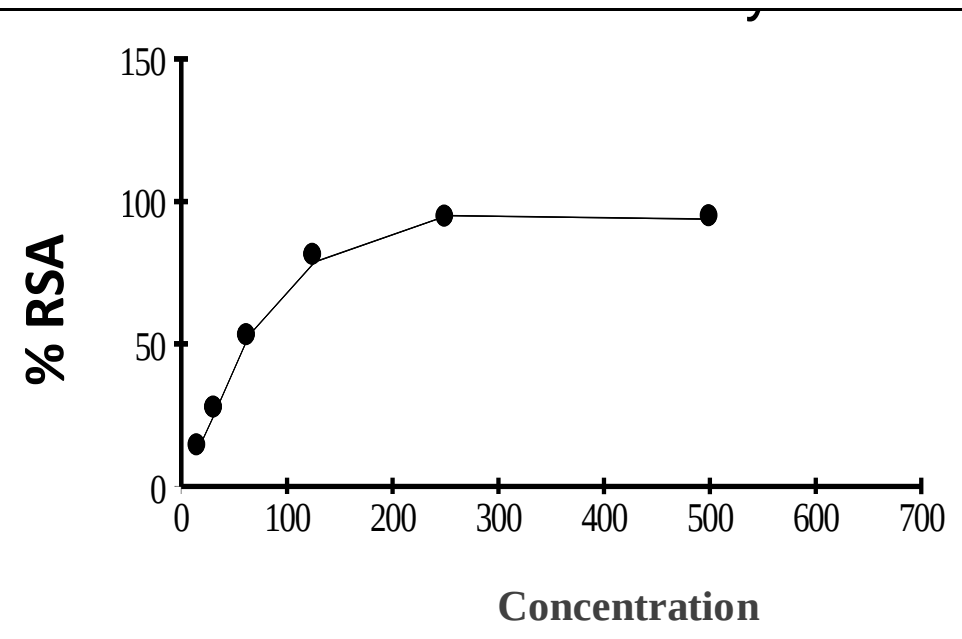

Comp 1 8

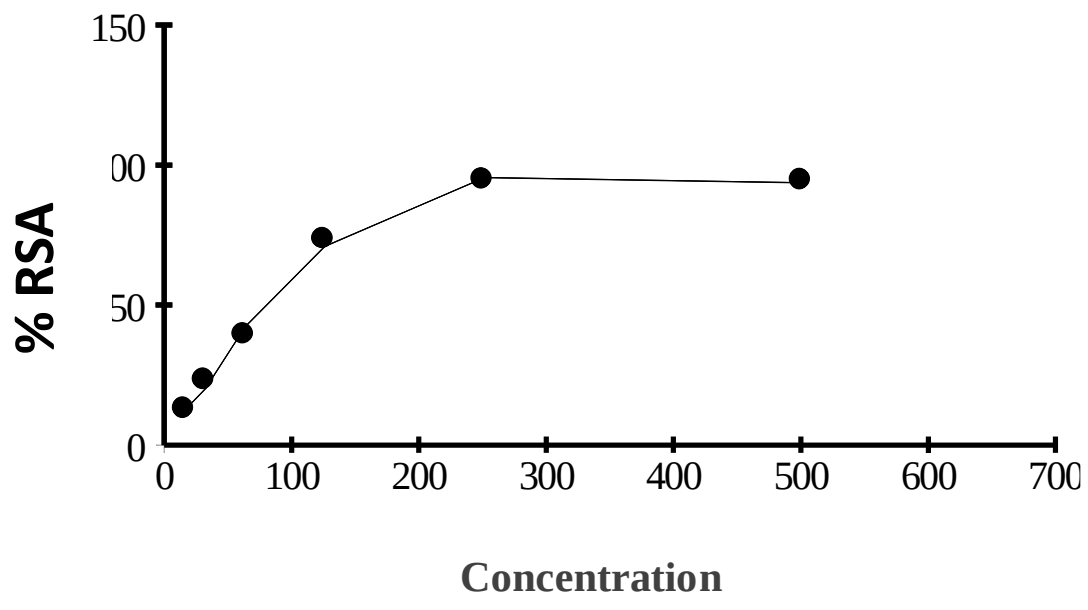

Comp 3

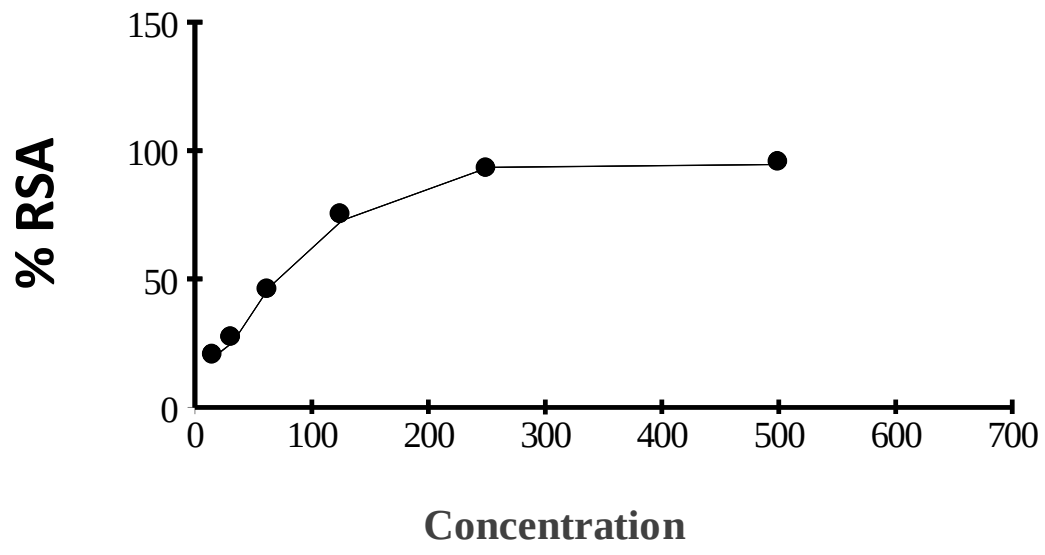

Comp 1 4

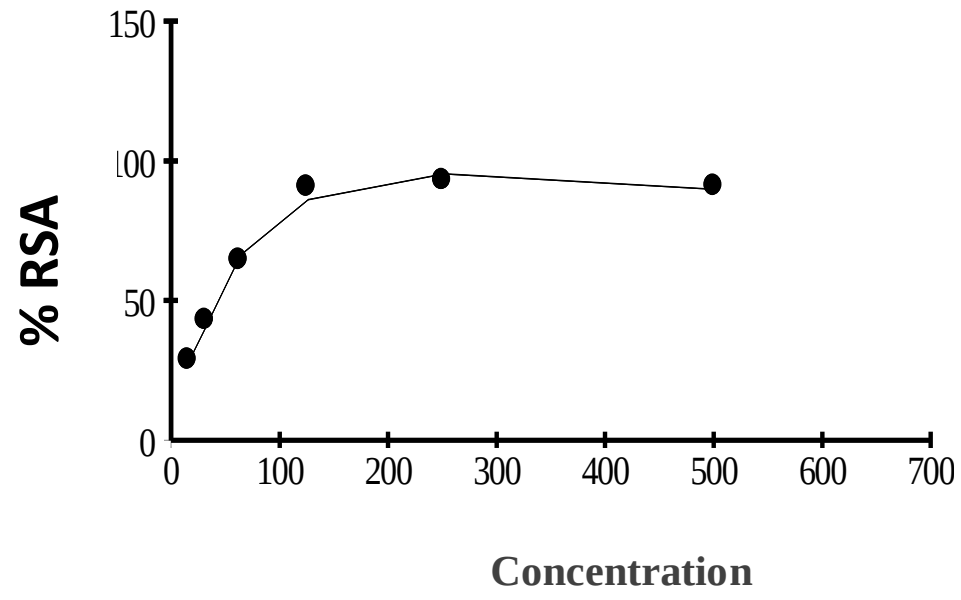

Comp 1 1

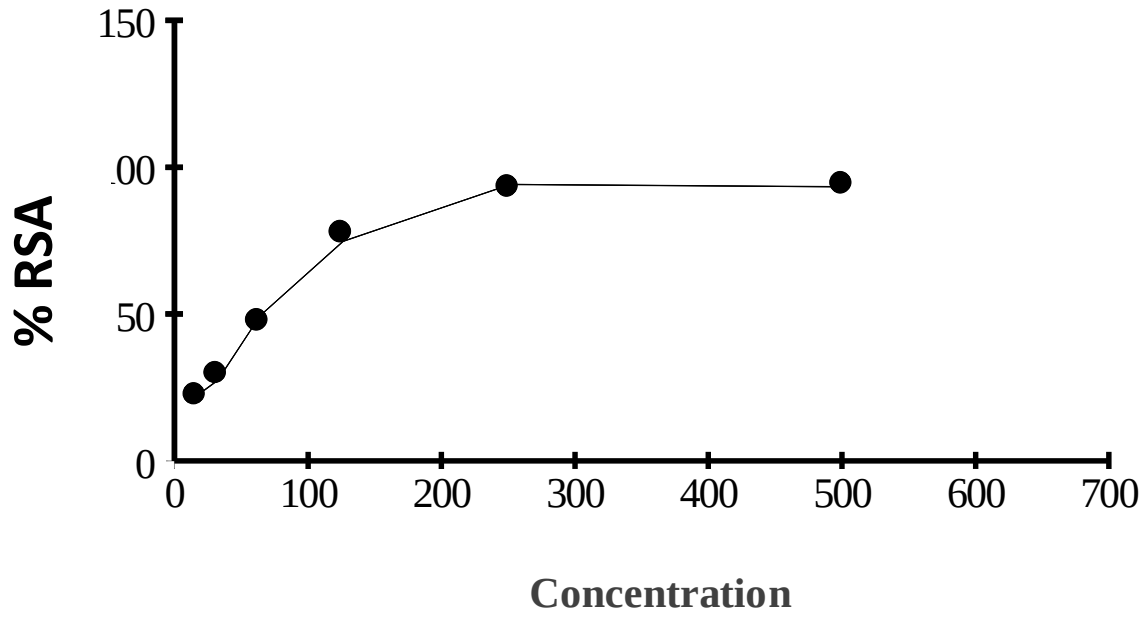

Comp 6

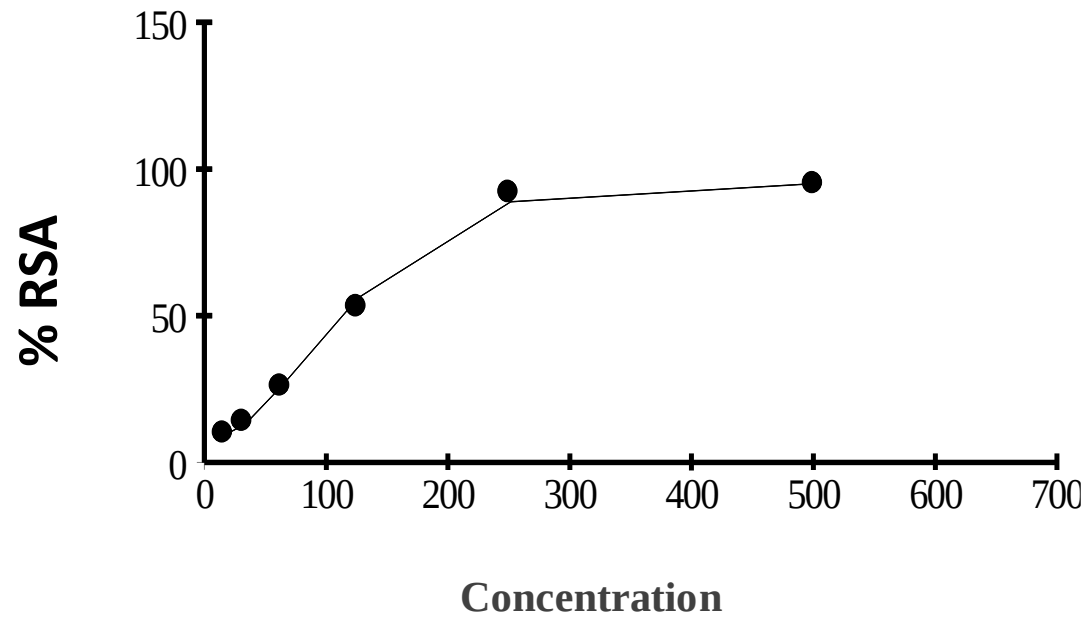

Comp 1 3

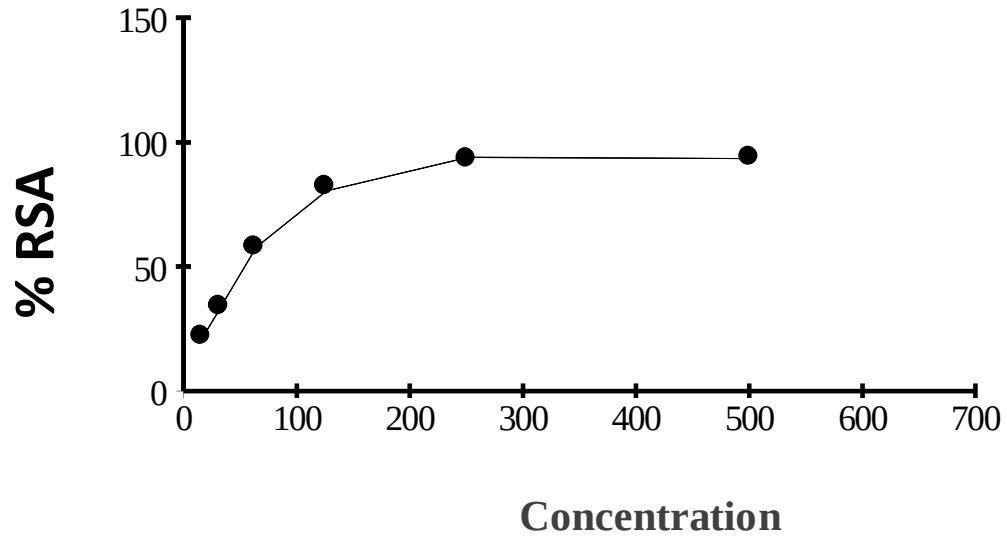

Comp 8

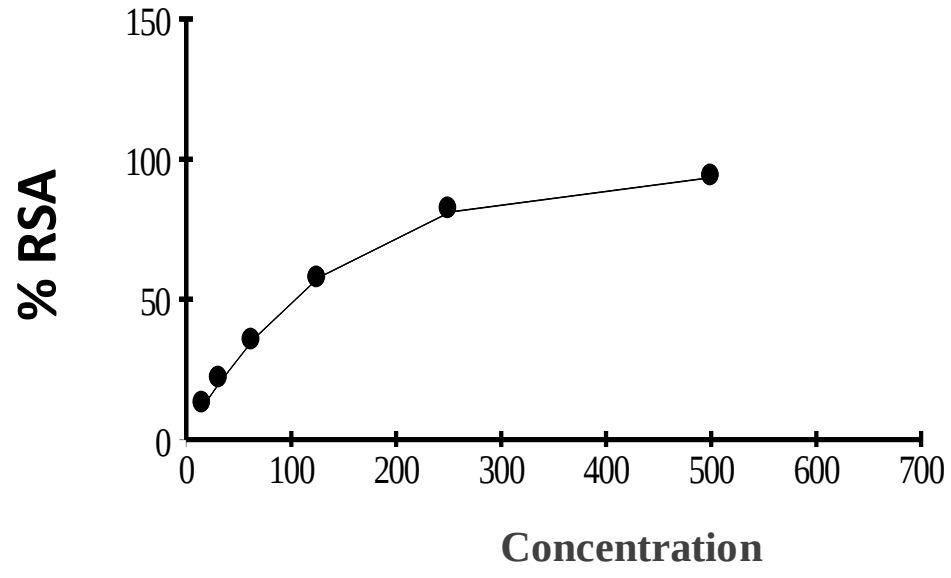

Comp 7

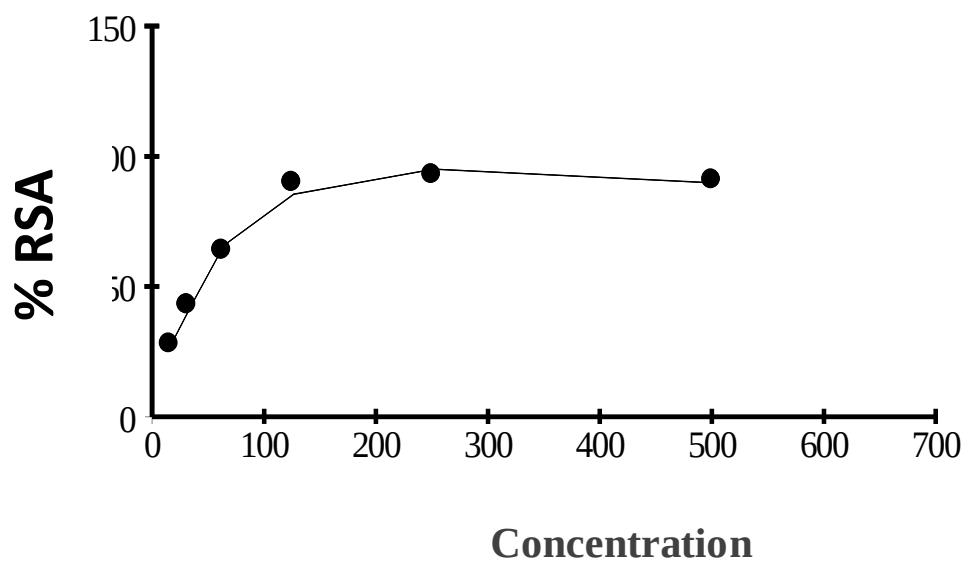

Comp 1 2

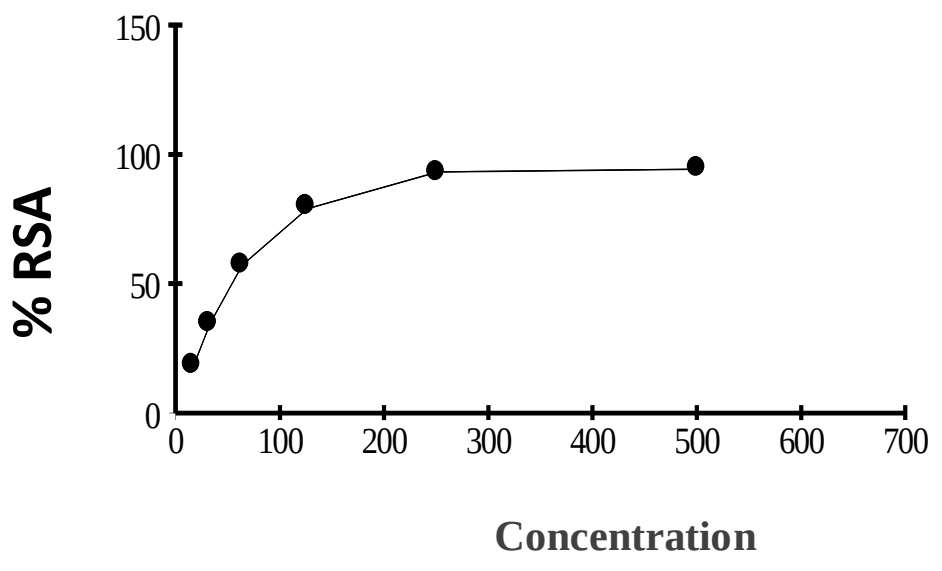

Comp 2

**Figure.1:** Thiocarbohydrazone derivatives (1-18) scavenge free radical in concentrations dependent manner at 517nm absorbance. Each graph plotted with % RSA at Y-axis and increasing concentration of compound at X-axis.

### *3.3 Structure Activity Relationship*

Structure activity relationship (SAR) relationship of THCs derivatives **1-18** (Table 2 and 2) established. Based on the observation, substitution of $-CH_3$ at C-4 position gives potent inhibitory activity of comp **10** with $IC_{50}$ = 24.2 ± 0.12 µM as compared to (Compound 4) 40.4 ± 1.78 µM at C-4 position however activity decreases when two $-CH_3$ groups added to C-2,3 position of compound **17** having 74.3 ± 1.49 µM.

Furthermore, the position of halogen substitution on benzyl ring also played an important role in scavenging activity. Substitution of Cl on compound 2, were found to have good activity than BHT with ($IC_{50}$ = 51.4 ±1.3µM, however radical scavenging activity increase with the addition of one more $CH_3$ on phenyl ring 76.6 ±1.7 µM and 51.5 ± 0.7 µM) respectively. Among different halogens substitution at C-4 position and F substitution at C-4 position (Compound 16) reduced the $IC_{50}$ the most to 38.4 ± 0.59 µM compared to $IC_{50}$ for –Br substitution on compound 15, 60.6 ± 2.47 µM at C-2 position. $NO_2$ substitution at C-4 position of benzylidine indicated that C-4 substitution gives the moderate inhibitory activity with $IC_{50}$ = 103.0 ± 2.73 µM in case of Compound 7, while, the substitution of – $NO_2$ at C-2, resulted a significant activity (Compound 18) with $IC_{50}$ = 59.3 ± 0.54 µM comparable to BHT. However, ethoxy substitution (Compound 1) at C-4 dramatically increased the $IC_{50}$ with 36.5 ± 1.7 µM as compare with compound 14 $IC_{50}$ = 68.3 ± 0.28 µM at C-2 position.

Substitution of indol and thio groups at C-1 and C-4 positions of compounds 9 and 12 gives significant activities $IC_{50}$ = 36.9 ± 0.64 µM and 40.4 ± 1.78 µM as compare to substitution of isopropyl and pyridine groups at C-4 positions of compounds 11 and 13 with $IC_{50}$ = 208.5 ± 3.93 µM and 109.3 ± 0.98 µM. on the other hand, anthracen group at C-9 positions of compounds 6 and un substituted on benzylidine ring of (Compounds 5) gives $IC_{50s}$ = 61.5 ± 0.83 µM and 42.4 ± 0.54 µM respectively.

## Table-1: Thiocarbohydrazones derivatives (1-18)

| Compound | Structure | IUPAC name | M.W (a.m.u) |
|---|---|---|---|
| **1** | 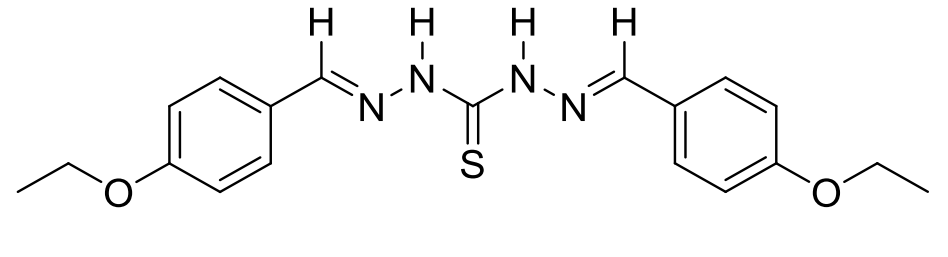 | *N,N-bis*[(E)-(4-Ethoxyphenyl)methylidene]carbothioic dihydrazide | 370.47 |
| **2** | 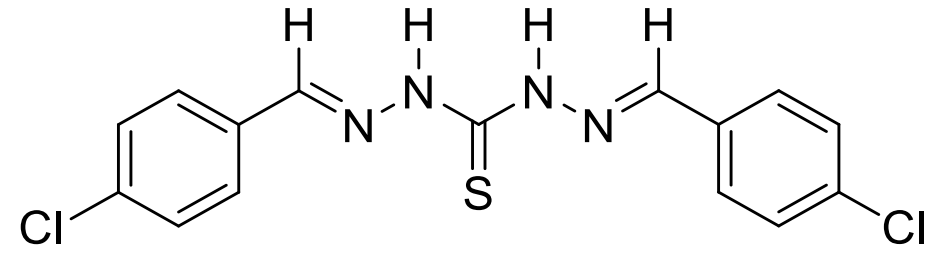 | *N,N-bis*[(E)-(4-Chloro)methylidene]carbothioic dihydrazide | 351.25 |
| **3** | 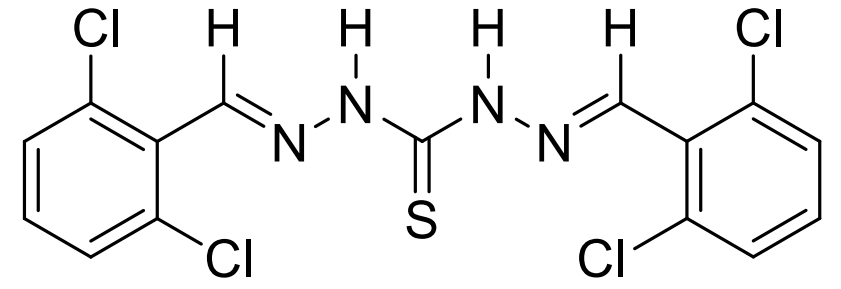 | *N,N-bis*[(E)-(2,6-Dichlorophenyl)methylidene]carbothioic dihydrazide | 420.14 |
| **4** | 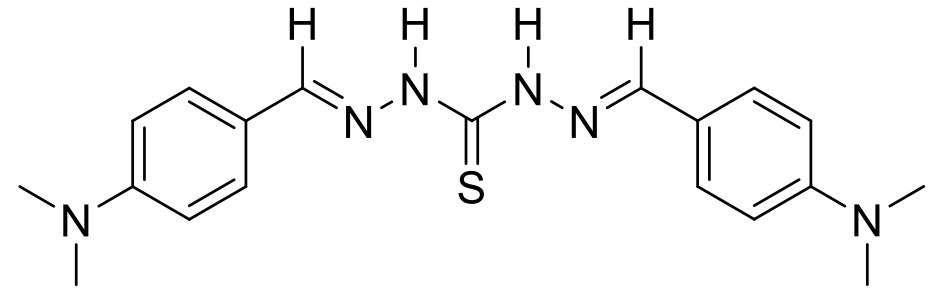 | *N,N*-bis[(E)-(4-Dimethylamino)methylidene]carbothioic dihydrazide | 368.50 |
| **5** | 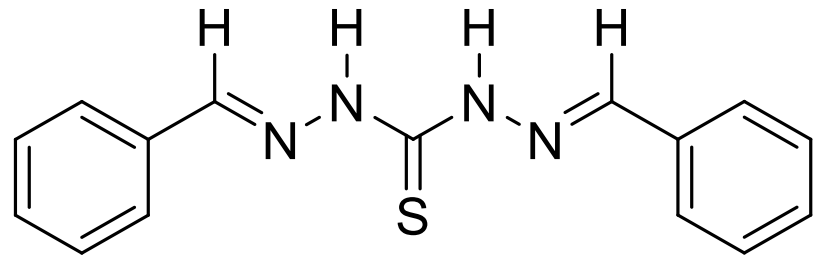 | *N,N-bis*[(E)Phenylmethylidene]carbothioic dihydrazide | 282.36 |
| **6** | 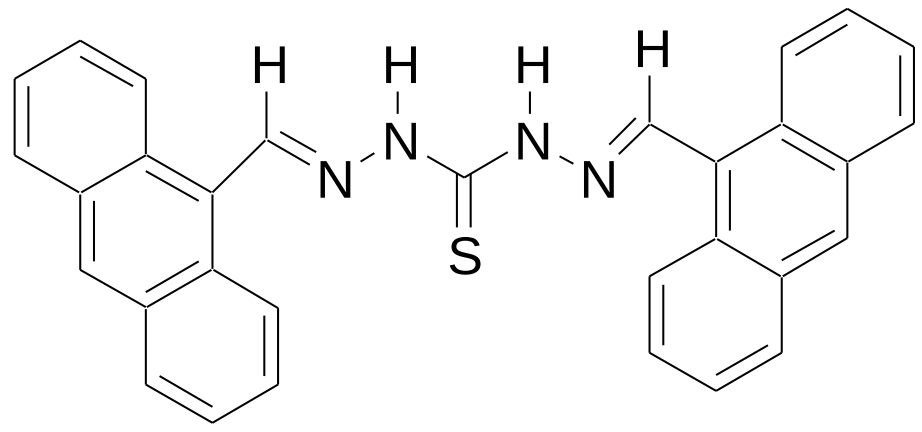 | *N,N-bis*[(E)-(9-anthryl Methylidene]carbothioic dihydrazide | 482.59 |
| **7** | 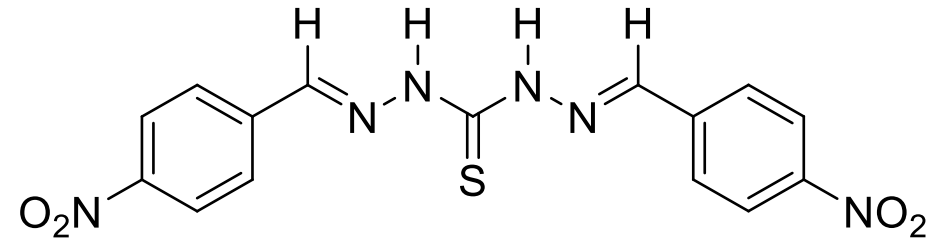 | *N,N-bis*[(E)-(4-Nitrophenyl)methylidene]carbothioic dihydrazide | 372.36 |
| **8** | 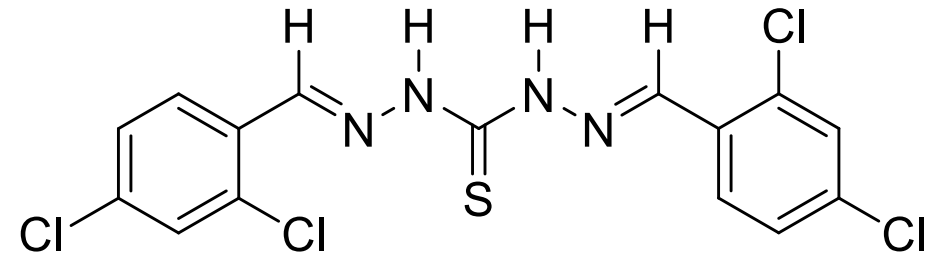 | *N,N-bis*[(E)-(2,4-Dichlorophenyl)methylidene]carbothioic dihydrazide | 420.14 |
| **9** | 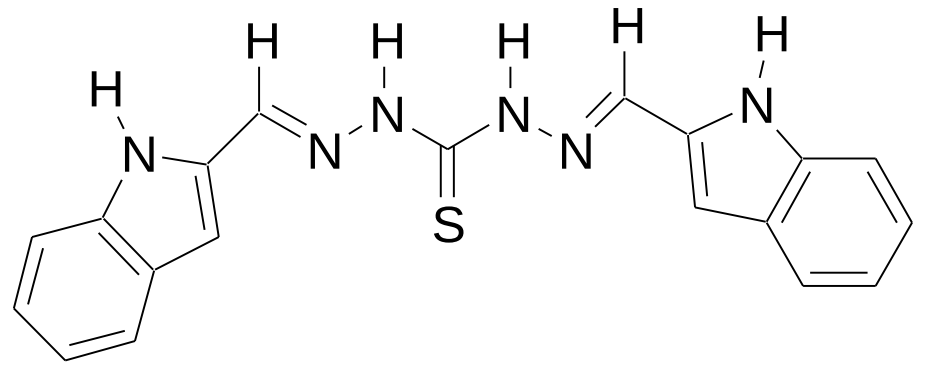 | *N,N-bis[*(E)-(1H Indol-2-yl Methylidene]carbothioic dihydrazide | 360.4 |
| **10** | 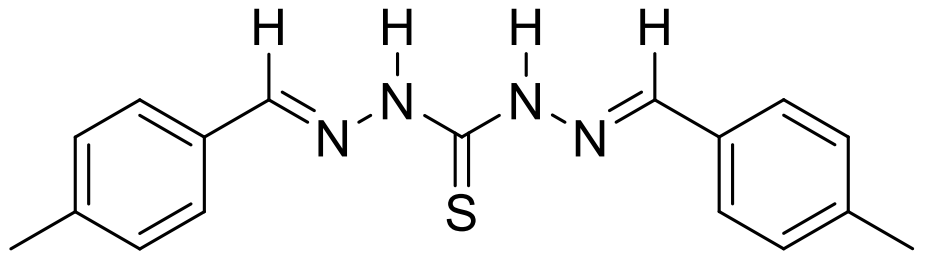 | *N,N-bis*[(E)-(4-Methylphenyl)methylidene]carbothioic dihydrazide | 310.42 |
| **11** | 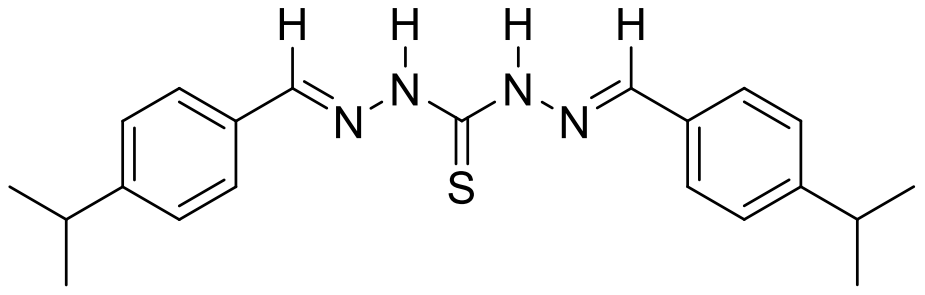 | *N,N-bis*[(E)-(4-Ispropylphenyl)methylidene]carbothioic dihydrazide | 366.52 |
| **12** | 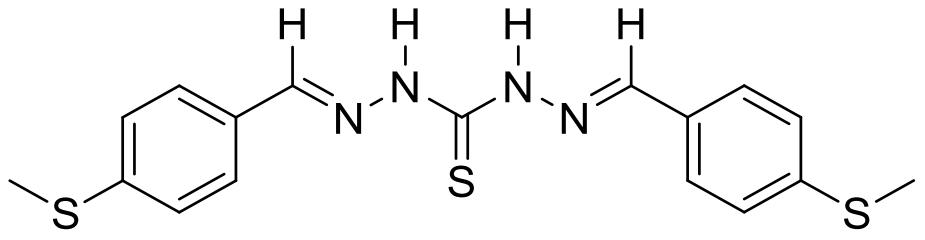 | *N,N-bis*{(E)-[ 4-(Methylsulfanyl)methylidene]carbothioic dihydrazide | 374.55 |
| **13** | 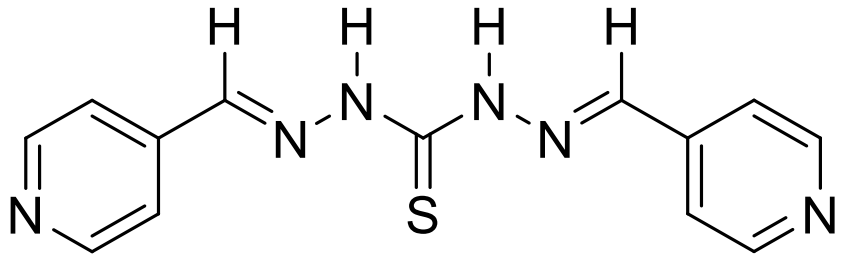 | *N,N-bis*[(E)- 4-Pyridinylmethylidene]carbothioic dihydrazide | 284.34 |
| **14** | 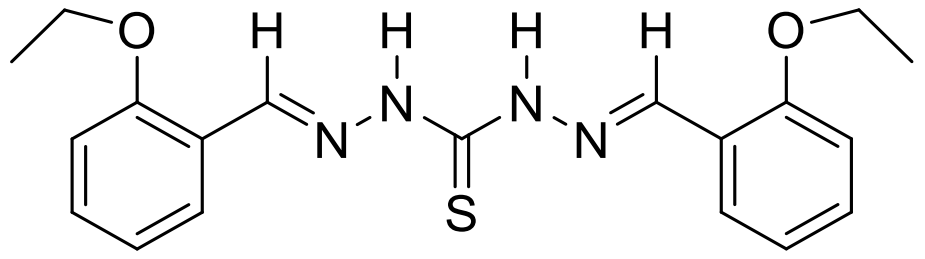 | *N,N-bis*[(E)-(2-Ethoxyphenyl)methylidene]carbothioic dihydrazide | 370.47 |
| **15** | 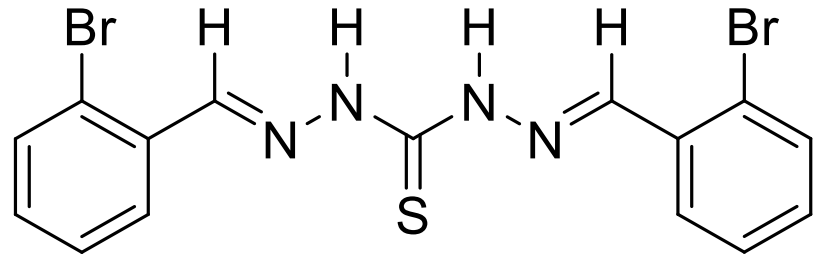 | *N,N-bis*[(E)-(2-Bromophenyl)methylidene]carbothioic dihydrazide | 440.16 |
| **16** | 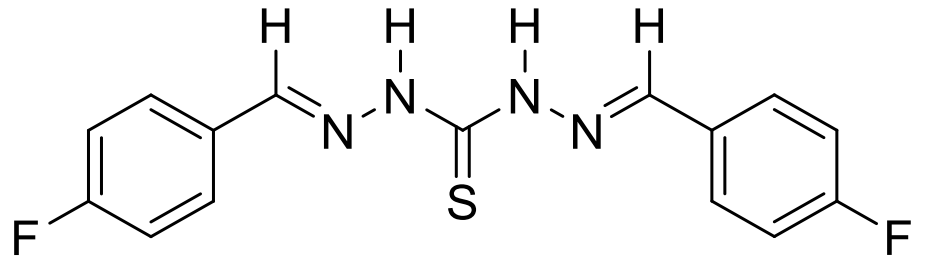 | *N,N-bis*[(E)-(4-Florophenyl)methylidene]carbothioic dihydrazide | 318.34 |
| **17** | 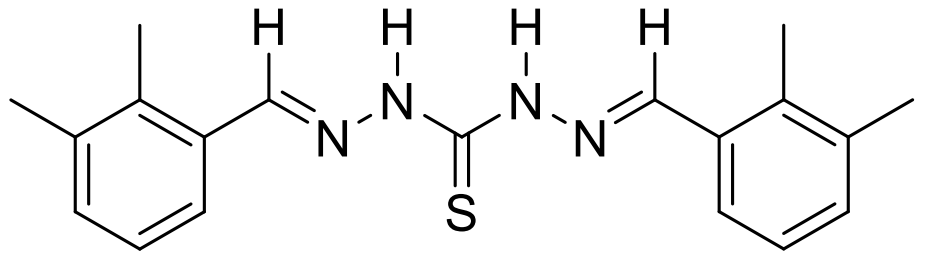 | *N,N-bis*[(E)-(2,3-Dimethylphenyl)methylidene]carbothioic dihydrazide | 338.47 |
| **18** | 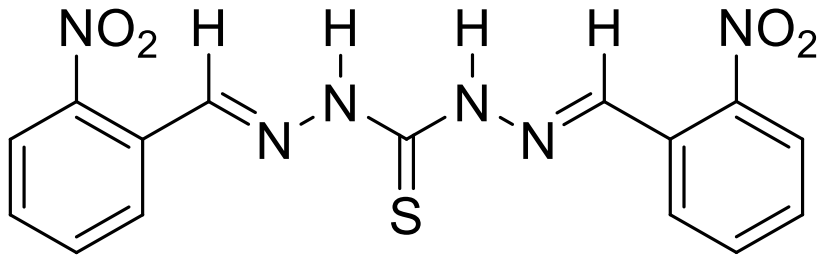 | *N,N-bis*[(E)-(2-Nitrophenyl)methylidene]carbothioic dihydrazide | 372.36 |

M.W: Molecular weight

**Table-2: Free Radical Scavenging Activities of compounds (1-18)**

| Compounds | % RSA | $IC_{50}$ ± SEM (μM) | Compounds | % RSA | $IC_{50}$ ± SEM (μM) |
|---|---|---|---|---|---|
| **1** | 88.886 | 36.5 ± 1.78 | **10** | 94.233 | 24.21 ± 0.12 |
| **2** | 95.019 | 51.4 ± 1.31 | **11** | 90.686 | 208.5 ± 3.93 |
| **3** | 94.613 | 76.6 ± 1.71 | **12** | 90.725 | 41.6 ± 0.98 |
| **4** | 81.407 | 40.41 ± 1.78 | **13** | 96.488 | 109.3 ± 0.98 |
| **5** | 95.150 | 42.4 ± 0.54 | **14** | 95.576 | 68.3 ± 0.28 |
| **6** | 94.761 | 61.5 ± 0.83 | **15** | 95.554 | 60.63 ± 2.47 |
| **7** | 93.971 | 103.0 ± 2.73 | **16** | 95.139 | 38.4 ± 0.59 |
| **8** | 95.019 | 51.5 ± 0.76 | **17** | 95.477 | 74.3 ± 1.49 |
| **9** | 89.824 | 36.9 ± 0.64 | **18** | 93.431 | 59.3 ± 0.54 |
| **BHT** | % | **128.8 ± 2.1** | | | |

RSA: % Radical scavenging activity

### *3.3. Cytotoxicity Against 3T3 Fibroblast Cells*

The cytotoxicity of compounds **1-18** were carried out by using mouse fibroblast 3T3 cell line. Derivatives **1**, **2, 4, 5, 6, 7, 10, 11, 13, 16** and **17** exhibited non-cytotoxic with $IC_{50}$ values ≥ 30 AS presented in Table 3. Derivatives **3, 9, 12** and **18** were found to have moderate cytotoxic effect with $IC_{50}$ values ranging from 10.35 ± 1.95 to 19.807 ± 1.95 µM, respectively. However, compound 8, 14 and 15 were found to be cytotoxic with $IC_{50}$ value of 6.45 ± 0.037 to 27.414 ± 0.037µM.

Table-3: Cytotoxicity studies of selected pyrimidine derivatives compounds (**1-18**)

| **Compound** | **% Inhibition** | **$IC_{50}$ ± SEM (µM)** | **Compound** | **% Inhibition** | **$IC_{50}$ ± SEM (µM)** |
|---|---|---|---|---|---|
| **1$^{o}$** | Non cytotoxic | ≥30 | **10$^{o}$** | Non cytotoxic | ≥30 |
| **2$^{o}$** | Non cytotoxic | ≥30 | **11$^{o}$** | Non cytotoxic | ≥30 |
| **3*** | Moderately Cytotoxic | 13.471 ± 0.812 | **12*** | Moderately Cytotoxic | 19.807 ±1.95 |
| **4$^{o}$** | Non cytotoxic | ≥30 | **13$^{o}$** | Non cytotoxic | ≥30 |
| **5$^{o}$** | Non cytotoxic | ≥30 | **14^** | Cytotoxic | 6.45 ±0.037 |
| **6$^{o}$** | Non cytotoxic | ≥30 | **15^** | Cytotoxic | 27.414 ±0.756 |
| **7$^{o}$** | Non cytotoxic | ≥30 | **16$^{o}$** | Non cytotoxic | ≥30 |
| **8^** | Cytotoxic | 24.103± 1.032 | **17$^{o}$** | Non cytotoxic | ≥30 |
| **9*** | Moderately Cytotoxic | 10.856± 0.839 | **18*** | Moderately Cytotoxic | 11.395 ±0.246 |
| **Cyclohexamide** | Cytotoxic | | | 0.26 ± 0.1 | |

$^{o}$ (Non cytotoxic), *(Moderately Cytotoxic), ^(Cytotoxic)

## 4. Discussion

The present study investigated an *in vitro* free radical scavenging activity of THCs (**1-18**) derivatives by employing DPPH assay our results that revealed that all (**1-18**) exhibit radical scavenging activity that is influenced apparently based on nature and position of substituents. Several THCs derivatives have been previously reported to possess higher antioxidant activity [27]. In this studied derivative **10** was found to be the most potent inhibitor of the series with ($IC_{50}$ = 24.2 ± 0.12 µM) as compared to BHT ($IC_{50}$ = 128.8 ± 2.1 µM) in addition with its non-cytotoxic nature towards normal cells. Whereas all other compounds of the series also exhibited many folds greater activity compared to the BHT ($IC_{50}$ = 128.8 ± 2.1 µM). Therefore, our future prospective work aims at investigating the *in vivo* free radical scavenging activities of the non-cytotoxic TCHs derivatives of the series by employing oxidative stress induced cell model that may result in clinically useful free radical scavengers to develop drug against oxidative stress induced cancer.

## Conflict of Interest

We declare that we have no conflict of interest.

## Acknowledgments

The authors are thankful to Molecular Bank of Dr. Panjwani Center for Molecular Medicine and Drug Research, International Center for Chemical and Biological Sciences, University of Karachi to provide studied compounds.